\ifdefined\TWOCOL
\documentclass[journal]{IEEEtran}
\else
\documentclass[journal,draftcls,onecolumn,12pt,twoside]{IEEEtran}
\fi

\usepackage[utf8]{inputenc} 
\usepackage{amsmath,amsfonts}
\usepackage{url}
\usepackage{cite}
\usepackage{bm}

\usepackage{pgfplots}
\pgfplotsset{compat=1.14}
\usepackage{tikz}
\usetikzlibrary{arrows, shapes}
\usepackage{mathtools}

\newcommand{\Fig}[1]{Fig.~\ref{#1}}

\newtheorem{theorem}{Theorem}
\newtheorem{lemma}{Lemma}

\newtheorem{define}{Definition}

\newtheorem{remark}{Remark}

\DeclareMathOperator*{\argmax}{arg\,max}
\DeclareMathOperator*{\argmin}{arg\,min}

\newcommand{\norm}[1]{\left\| #1 \right\|}
\newcommand{\normf}[1]{\norm{#1}_F^2}
\newcommand{\Pb}[1]{\Pr\left[#1 \right]}

\newcommand{\Exp}[1]{\exp\left(#1 \right)}
\newcommand{\determ}[1]{\left|#1\right|}

\newcommand\cn[1]{\mathcal{CN}\left(#1\right)}

\newcommand\Span[1]{span\left\{#1\right\}}
\newcommand\rscalar[1]{\mathsf{#1}}
\newcommand\rvec[1]{\mathsf{\mathbf{#1}}}
\newcommand\rmat[1]{\bm{\mathsf{#1}}}
\newcommand\dvec[1]{\bm{#1}}
\newcommand\dset[1]{\mathcal{#1}}

\newcommand\dmat[1]{\bm{#1}}

\begin{document}

\title{Unsourced Random Access with the MIMO Receiver: Projection Decoding Analysis}

\author{Kirill Andreev\IEEEauthorrefmark{1}\thanks{\IEEEauthorrefmark{1}Kirill Andreev, Daria Ustinova and Alexey Frolov are with the Center for Next Generation Wireless and IoT (NGW), Skolkovo Institute of Science and Technology, Moscow, Russia (emails: k.andreev@skoltech.ru, d.ustinova@skoltech.ru, al.frolov@skoltech.ru).}
\hspace{0.7cm} \and Daria Ustinova\IEEEauthorrefmark{1}
\hspace{0.7cm}\and Alexey Frolov\IEEEauthorrefmark{1}
\thanks{The research was carried at Skolkovo Institute of Science and Technology and supported by the Russian Science Foundation (project no. 18-19-00673), \protect\url{https://rscf.ru/en/project/18-19-00673/}}
}

\maketitle

\begin{abstract}

We consider unsourced random access with MIMO receiver -- a crucial communication scenario for future 5G/6G wireless networks. We perform a projection-based decoder analysis and derive energy efficiency achievability bounds when channel state information is unknown at transmitters and the receiver (no-CSI scenario). The comparison to the maximum-likelihood (ML) achievability bounds by Gao et al. (2023) is performed. We show that there is a region where the new bound outperforms the ML bound. The latter fact should not surprise the reader as both decoding criteria are suboptimal when considering per-user probability of error (PUPE). Moreover, transition to projection decoding allows for significant dimensionality reduction, which greatly reduces the computation time.

\end{abstract}

\begin{IEEEkeywords}
Energy efficiency, fundamental limits, finite blocklength, unsourced random access, MIMO
\end{IEEEkeywords}

\section{Introduction}

Massive machine-type communications (mMTC) is a new communication scenario for 5G/6G wireless networks featuring battery-powered devices sending short packets sporadically. The main challenges of mMTC scenario such as large amount of devices, short packets and partial user activity make grant-based communications highly inefficient. We also note that the use of different encoders (or different codebooks) leads to a large receiver complexity as the decoder should first detect which encoders were utilized. Thus, the most promising strategy is to use grant-free access (or random access) with the same codebook. This kind of random access is known in the literature as  \textit{unsourced random access} (URA) as it decouples identification and decoding tasks. The formal statement of the URA problem is proposed in~\cite{polyanskiy2017perspective}. The fundamental limits and low-complexity schemes for the Gaussian multiple access channel (MAC) are presented in \cite{polyanskiy2017perspective, ZPT-isit19, achRandUser2023, vem2017user, Marshakov2019Polar, codedCS2020, fengler2020sparcs, AmalladinneAMP2021, AmalladinneJoint2023, Duman2021, PradhanLDPCSC2021}. More realistic fading channel models were also considered in the literature. The papers~\cite{Kowshik2021TIT, Kowshik2020TCOM, Truhachev2022, Andreev2022} describe fundamental limits and practical schemes for the single-antenna quasi-static Rayleigh fading MAC. The authors of~\cite{Fengler2021} were the first who observed that the use of multiple antennas at the base station can significantly improve energy efficiency and do not affect the  complexity of transmitters. Recall that the transmitters are autonomous battery-powered devices and should be as simple as possible. Practical schemes for URA with MIMO receiver were widely considered in the literature. The design of practical schemes relies on pilot-based channel estimation followed by decoding and successive interference cancellation (SIC) \cite{Fengler2022, Duman2022} or coded compressed sensing approach in combination with non-Bayesian activity detection \cite{Fengler2021}, SIC \cite{AmalladinneMIMO2021}, or the use of slot-wise correlations \cite{Shyianov2020}. The scheme \cite{FASURA} combines the above mentioned approaches and provides state-of-the-art performance for the MIMO channel. 

It is important to understand the fundamental limits of the URA with the MIMO receiver to judge how good the proposed practical schemes are. These limits were unknown for a long time and appeared recently in~\cite{Poor2022ML} for the different codebook case and in~\cite{noPoorUnsourcedML2023} for the URA case. The papers provide converse and maximum likelihood (ML) achievability bounds for known and unknown CSI scenarios. In this paper, we continue this line of research and extend the results for the case of projection decoding. The projection decoding can work only when the number of active users is smaller than the frame length for any number of receive antennas. Hence, this method is inappropriate for obtaining scaling laws (as mentioned in~\cite{Poor2022ML}). At the same time, we show this method to provide good finite-length performance.

Our contribution is as follows. We derive energy efficiency achievability bounds for the URA with MIMO receiver utilizing the projection decoding. We analyze the results for $50$ and $64$ receive antennas and demonstrate that there is a region where the new bound outperforms the ML bound from~\cite{noPoorUnsourcedML2023}. The latter statement should not surprise the reader, as the ML decoder is optimal if we deal with the joint error. At the same time, the paper deals with per-user error, and both methods are suboptimal in this case. Moreover, the transition to projection decoding allows for significant dimensionality reduction and a great reduction in the computation time.

The paper has the following structure. The system model is presented in Section~\ref{sec:syst_model}. Section~\ref{sec:decoding_alg} is devoted to the decoding methods, namely maximum likelihood and projection-based decoding algorithms. Sections~\ref{sec:main_res} and ~\ref{sec:proof} are devoted to the main result statement and proof. In Section~\ref{sec:dem_reduction} we explain how to reduce the bound calculation time by means of the basis change. Finally, in Section~\ref{sec:numres}, we present the numerical results.

\textit{Notation:} Let $\mathbb{C}$ denote the set of complex numbers. We represent deterministic quantities with italic letters: scalar values are denoted by non-boldface letters (e.g., $x$ or $X$), vectors by boldface small letters (e.g., $\dvec{x}$) and matrices by boldface capital letters (e.g., $\dmat{X}$). Random quantities are denoted with non-italic letters with sans-serif font, e.g., scalar $\rscalar{x}$ or $\rscalar{X}$, vector $\bm{\mathsf{x}}$, and matrix $\rmat{X}$. We denote sets by calligraphic letters (e.g., $\dset{S}$). The letter $\dset{E}$ is reserved for events, by $\dset{E}^c$ we denote the complementary event to $\dset{E}$. For any positive integer $n$, we use the notation $[n] = \{1,\dots,n\}$. The $n \times n$ identity matrix is denoted by $\dmat{I}_n$. By $\determ{\dmat{A}}$ be denote the determinant of the matrix $\dmat{A}$. $\dmat{A}^H$ is a conjugate transpose of $\dmat{A}$. Let $\dset{S} = \{i_1, \ldots, i_s\} \subseteq [n]$ with $i_1 < \ldots < i_s$. Given the matrix $\dmat{A} = [\dvec{a}_1, \ldots, \dvec{a}_n]$, the restriction $\dmat{A}_\dset{S}$ of $\dmat{A}$ to $\dset{S}$ is the matrix $\dmat{A}_\dset{S} = [\dvec{a}_{i_1},\ldots,\dvec{a}_{i_s}]$. The letter $\dmat{P}$ is reserved for the orthogonal projection operator. By $\dmat{P}_{\dmat{X}}$ we mean the orthogonal projection\footnote{Recall that $\dmat{P}_{\dmat{X}} = \dmat{X}\left(\dmat{X}^H\dmat{X}\right)^{-1}\dmat{X}^H$ when the columns of $\dmat{X}$ are linearly independent, which holds for random Gaussian codebook} onto the subspace spanned by the columns of $\dmat{X}$. We use the notation $\dmat{P}_{\dset{S}} = \dmat{P}_{\dmat{X}_\dset{S}}$ in the cases when the matrix $\dmat{X}$ is clear from the context, and $\dmat{P}_{\dmat{X}}^{\perp} = \dmat{I} - \dmat{P}_{\dmat{X}}$. By $\norm{\dvec{x}}_2$ and $\norm{\dmat{X}}_F$ we denote the Euclidean norm of the vector $\dvec{x}$ and the Frobenius norm of the matrix $\dmat{X}$. By $\mathcal{N}\left( \bm{0},\dmat{I}_n \right)$ and $\cn{\bm{0},\dmat{I}_n}$ we denote standard and complex standard normal random vectors. $\mathbb{E}$ denotes the expectation operator, and $\Pb{\dset{E}}$ means a probability of the event $\dset{E}$.

\section{System model}\label{sec:syst_model}

Consider the model presented in \cite{polyanskiy2017perspective}. Assume we have $K_\text{tot} \gg 1$ users in the system, and only $K_a \ll K_\text{tot}$ of them are active at any time instance. We assume the number of active users ($K_a$) to be known at the receiver. Communication takes place in a frame-synchronized manner. Each frame has a length of $n$ complex channel uses. All users utilize one message set $[M]$ and the same encoder function $f: [M] \to \mathbb{C}^n$. A natural power constraint $||f(W)||^2_2 \leq nP, W \in [M]$, is also required.

In what follows we consider the static Rayleigh fading channel and the base station equipped with $L \geq 2$ receive antennas. Each device has the only transmit antenna to increase the battery life. Let $\dset{T} = \{W_1, W_2, \ldots, W_{K_a}\}$ be the set of transmitted messages, the channel model can be described as follows
\begin{equation}
\label{eq:channel_model}
\rmat{Y} = \dmat{X} {\dmat{\Phi}}(\dset{T}) \rmat{H} + \rmat{Z},
\end{equation}
where $\dmat{Y} = [\dvec{y}_1, \ldots, \dvec{y}_L] \in \mathbb{C}^{n \times L}$ is the channel output matrix, $\dmat{X} = [\dvec{x}_1 \ldots \dvec{x}_M] \in \mathbb{C}^{n \times M}$, where $\dvec{x}_W = f(W), W \in [M]$ is the common codebook, $\dmat{\Phi}(\dset{T}) \in \{0,1\}^{M \times K_a}$ is the activity matrix (each column has the only unit element in the position corresponding to the user's message), $\mathbf{Z} \in \mathbb{C}^{n \times L}$ is the Gaussian noise matrix, each elements is sampled i.i.d from $\cn{0,1}$, and finally $\mathbf{H} \in \mathbb{C}^{K_a \times L}$ is the matrix of fading coefficients, each element is sampled i.i.d from $\cn{0,1}$. Note that fading coefficients are independent of codewords and $\dmat{Z}$. In what follows we assume $\dmat{H}$ to be unknown at
transmitters and the receiver (unknown CSI scenario). 

Decoding is performed up to permutation\footnote{Note that the channel~(\ref{eq:channel_model}) is permutation invariant, thus we can perform such type of decoding.}. We require the decoder to produce a set ${\dset{R}} \subseteq [M]$. In what follows we assume the receiver to know the number of active users $K_a$, thus we require $| {\dset{R}} | = K_a$. The main performance measure we use is the Per User Probability of Error (PUPE)
\[
P_e = \frac{1}{K_a} \sum\limits_{i=1}^{K_a} \Pb{W_i \not\in {\dset{R}}},
\]
note that we do not need to consider False Alarm Rate (FAR) in this case as the output size is fixed. Our goal is to minimize the energy-per-bit ($E_b/N_0 = Pn/k$, $k=\log_2M$) spent by each user.

\section{Decoding algorithms}\label{sec:decoding_alg}

Note that $\Pb{\bigcup\nolimits_{i \ne j} \{ W_i = W_j \}} \leq \binom{K_a}{2}/M$ which is negligible for the system parameters of interest. Thus in what follows we do not consider the case of colliding messages and require the decoder to return a set ${\dset{R}} \subseteq [M], |{\dset{R}}| = K_a$. Due to this fact in what follows we replace $\dmat{X}{\dmat{\Phi}}(\dset{T})$ with $\dmat{X}_{\dset{T}}$.

Let us start with the \textit{maximum likelihood (ML) decoding rule} from \cite{Poor2022ML, noPoorUnsourcedML2023}:
\begin{equation}\label{eq:ml_decoder}
\dset{R} = \argmax_{\dset{R}' \subset [M], \left| \dset{R}'\right| = K_a} \Pb {\dmat{Y} \middle| \dmat{X}_{\dset{R}'} },
\end{equation}
where $\Pb {\dmat{Y} \middle| \dmat{X}_{\dset{R}'} } = \mathbb{E}_{\rmat{H}}\left\{\Pb {\dmat{Y} \middle| \dmat{X}_{\dset{R}'}, \rmat{H}}\right\}$.

In this work following the ideas of \cite{Kowshik2020TCOM} we consider \textit{projection decoding rule}:
\begin{equation}\label{eq:proj_decoder}
\dset{R} =  \argmax_{\dset{R}' \subseteq [M], \left|\dset{R}'\right| = K_a} \max_{\dmat{H}}\left\{ \Pb{\dmat{Y} \middle| \dmat{X}_{\dset{R}'}, \dmat{H}}\right\}.
\end{equation}

After some transformations we have

\begin{eqnarray}
    \label{eq:max_proj_problem}
    \dset{R} &=&  \argmin_{\dset{R}' \subseteq [M], |\dset{R}'| = K_a} \min_{\dmat{H}} \norm{\dmat{Y} - \dmat{X}_{\dset{R}'} \dmat{H}}_F^2 \nonumber \\
    &=& \argmin_{\dset{R}' \subseteq [M], |\dset{R}'| = K_a} \left(\norm{\dmat{Y}}^2_F - \norm{\dmat{P}_{\dset{R}'}\dmat{Y}}^2_F \right) \nonumber \\
    &=& \argmax_{\dset{R}' \subseteq [M], |\dset{R}'| = K_a} \norm{\dmat{P}_{\dset{R}'}\dmat{Y}}^2_F. \label{eq:dec_cond}
\end{eqnarray}

\section{Main result}\label{sec:main_res}

Let us introduce the definition of ensemble and formulate the main result.

\begin{define}
Let $\mathcal{E}(n, M, P)$ be the ensemble of codebooks $\dmat{X}$ of size $n \times M$, where each element is sampled  i.i.d. from $\cn{0,P}$.
\end{define}

\begin{theorem}\label{thm}
    Let $P' < P$. There exists a codebook $\dmat{X}^* \in \mathcal{E}(n, M, P')$ satisfying power constraint $P$ and providing
\begin{eqnarray*}
    P_e \leq  \sum_{t = 1}^{K_a} \frac{t}{K_a} \inf_{0 \leq \alpha \leq 1}(p_{1,t} + p_{2,t}) + p_0,
\end{eqnarray*}
with
\begin{equation}
p_0 =\frac{\binom{K_a}{2}}{M}+K_a\Pb{\frac{1}{2n}\sum_{i=1}^{2n}\xi_i^2> \frac{P}{P'}}, \:\: \xi_i \overset{i.i.d}{\sim} \mathcal{N}(0,1), \label{eq:p0}
\end{equation}
\begin{equation}
p_{1,t} = \binom{K_a}{t} \binom{M-K_a}{t} \mathbb{E}\left[ \inf_{u, v > 0, \lambda_{\mathcal{D}} > 0}\determ{\dmat{I} - \rmat{D} \rmat{\Sigma}}^{-L} \right], \label{eq:p1t}
\end{equation}
\begin{equation}
p_{2,t} = \binom{K_a}{t} \mathbb{E} \left[ \inf_{\delta > 0, \lambda_{\mathcal{B}} > 0}\determ{\dmat{I} - \rmat{B}\rmat{\Sigma}}^{-L} \right], \label{eq:p2t}
\end{equation}
where 
\begin{eqnarray}
\rmat{\Sigma} &=& \dmat{I}_n + \rmat{X}_{[K_a]} \rmat{X}^H_{[K_a]}, \nonumber \\
\rmat{D} &=& u \rmat{P}_{[K_a + t] \backslash [t]} - u \rmat{P}_{[K_a]} + \alpha v \rmat{P}_{[K_a] \backslash [t]}^\perp - v \rmat{P}_{[K_a]}^\perp, \label{eq:D}\\
\rmat{B} &=&  - \alpha \delta \rmat{P}_{[K_a] \backslash [t]}^\perp + \delta \rmat{P}_{[K_a]}^\perp \label{eq:B}
\end{eqnarray}
By $\lambda_{\mathcal{D}}$ and $\lambda_{\mathcal{B}}$ we mean minimum eigenvalues of $\dmat{I} - \rmat{D} \rmat{\Sigma}$ and $\dmat{I} - \rmat{B} \rmat{\Sigma}$ accordingly. The expectations are taken over $\rmat{X}_{[K_a + t]}$.
\end{theorem}

\section{Proof of the main result}\label{sec:proof}

\subsection{Required lemmas}

\begin{lemma}[Chernoff bound, \cite{cover2012elements}]\label{lemma:chernoff}
Let $\xi_1$ and $\xi_2$ be any random variables, then for any $u,v > 0$ the following bounds hold:
\[
\Pb{\chi_1 \geq 0} \leq \mathbb{E}_{\chi_1} \left[ \exp\left( u \chi_1 \right) \right]
\]
and
\[
\Pb{\chi_1 \geq 0, \chi_2 \geq 0} \leq \mathbb{E}_{\chi_1, \chi_2} \left[ \exp\left( u \chi_1 + v \chi_2 \right) \right].
\]
\end{lemma}

\begin{lemma}[{Quadratic form expectation,~\cite{mathai1992quadratic}}]\label{lemma:qf_inv}
Let $\rvec{x} \sim \mathcal{CN}\left(0, \dmat{\Sigma}\right)$, $\dmat{A}$ be the Hermitian matrix and $\dmat{I} -\dmat{A} \dmat{\Sigma}$ be a positive definite matrix, then
\[
\mathbb{E}_{\rvec{x}} \left[ \Exp{\rvec{x}^H \dmat{A} \rvec{x} } \right] = \determ{\dmat{I} -\dmat{A} \dmat{\Sigma}}^{-1}.
\]
\end{lemma}

\subsection{$P_e$ estimate and $t$-error events}

Recall that $\dset{T}$ and ${\dset{R}}$ are the sets of transmitted and received messages. Let $\dset{C} = \dset{T} \bigcap {\dset{R}}$, $\dset{M} = \dset{T} \backslash {\dset{R}}$ and $\dset{F} = {\dset{R}} \backslash \dset{T}$ be the sets of correctly received, missed, and falsely detected messages accordingly.

Following the approach of \cite{polyanskiy2017perspective}, we consider $t$-error events $\dset{E}_t = \{|\dset{M}| = t\}$ and use the following estimate
\begin{eqnarray*}
P_e &\leq& \sum_{t = 1}^{K_a} \frac{t}{K_a} \Pb{\dset{E}_t} \\
    &+& \Pb{\bigcup\nolimits_{i \ne j} \{ W_i = W_j \}} \\
    &+& \Pb{\bigcup\nolimits_{i \in [K_a]} \{ \norm{\rvec{x}_{W_i}}^2 > Pn \} }.
\end{eqnarray*}

Clearly, the last two terms can be upper bounded by $p_0$ (see \eqref{eq:p0}). Thus, in what follows we do not consider colliding messages and power violation. W.l.o.g. assume $\dset{T} = [K_a]$. Note that
\[
\dset{E}_t = \bigcup_{\dset{M} \subseteq [K_a], \dset{F} \subseteq [M] \backslash [K_a]} \dset{E}(\dset{M}, \dset{F}),
\]
where (recall the decoding condition \eqref{eq:dec_cond})
\[
\dset{E}(\dset{M}, \dset{F}) = \left\{ \norm{\rmat{P}_{\dset{F} \bigcup \dset{C}}\rmat{Y}}^2_F \geq \norm{\rmat{P}_{[K_a]}\rmat{Y}}^2_F \right\},
\]
where we used the condition $\dset{M} \bigcup \dset{C} = \dset{T} = [K_a]$.

\subsection{Fano's trick}

The standard approach to upper bound $\Pb{\dset{E}_t}$ is to apply the union bound. This approach is known to overestimate the resulting probability and leads to bad results, thus following \cite{Poor2022ML, noPoorUnsourcedML2023} we apply the Fano's trick. Let us choose so-called ``good'' region $\dset{B}$, we proceed as follows 
\[
\Pb{\dset{E}_t} \leq \Pb{\dset{E}_t \bigcap \dset{B}} + \Pb{\dset{B}^c}.
\]

Let us specify the chosen ``good'' region. Following the ideas\footnote{Note that the papers are devoted to a single antenna case.} of \cite{Kowshik2021TIT, Kowshik2020TCOM}, we introduce the following region
\begin{equation}\label{eq:ball_prj}
\dset{B} = \bigcap\limits_{\dset{M} \subseteq [K_a]} \dset{B}_{\dset{M}},
\end{equation}
where
\[
\mathcal{B}_{\dset{M}} = \left\{\normf{\rmat{P}_{[K_a]}^\perp \rmat{Y}} \leq \alpha \normf{\rmat{P}_{\dset{C}}^\perp\rmat{Y}}\right\}.
\]
for some $0 \leq \alpha \leq 1$.

The intuition behind a good decoding region is the following. Consider a case when a projection onto orthogonal complement of the subspace spanned by transmitted codewords $P_{\dset{T}}^\perp$ has small Frobenius norm. In this case, it is likely that exists some codeword (not from the transmitted set) that is well-aligned with the received signal. Hence, the probability of error will be high.

\subsection{Estimating $\Pb{\dset{E}_t \bigcap \dset{B}}$ and $\Pb{ \dset{B}^c}$}

Let us start with $\Pb{\dset{E}_t \bigcap \dset{B}}$, we have
\begin{eqnarray*}
&&\Pb{\dset{E}_t \bigcap \dset{B}} \leq \Pb{ \bigcup_{\dset{M}, \dset{F}} \dset{E}(\dset{M}, \dset{F}) \bigcap \dset{B}_{\dset{M}}} \\
&=& \mathbb{E}_{\rmat{X}} \left[ \Pb{ \bigcup_{\dset{M}, \dset{F}} \dset{E}(\dset{M}, \dset{F}) \bigcap \dset{B}_{\dset{M}} \;\middle|\; \rmat{X} }  \right] \\
&\leq& \binom{K_a}{t} \binom{M-K_a}{t} \mathbb{E}_{\rmat{X}} \left[ \Pb{ \dset{E}(\dset{M}', \dset{F}') \bigcap \dset{B}_{\dset{M}'} \;\middle|\; \rmat{X} }  \right],
\end{eqnarray*}
where $\dset{M}' = [t]$, $\dset{F}' = [K_a + t]\backslash[t]$. The last transition is due to the union bound and symmetry property (as we are averaging over the codebook).

Let us consider $\tilde{p}_{1,t} = \Pb{ \dset{E}(\dset{M}', \dset{F}') \bigcap \dset{B}_{\dset{M}'} \;\middle|\; \rmat{X} = \dmat{X} }$. Note that the projection matrices are no more random in this case as they are functions of $\dmat{X}_{[K_a+t]}$.

Given $\dmat{X}_{[K_a]}$, the columns of $\rmat{Y}$ are independent and have the following distribution
$$
\rvec{y}_l \sim \mathcal{CN}(\mathbf{0}, \dmat{\Sigma}), \:\: l \in [L], \quad \dmat{\Sigma} = \dmat{I}_n + \dmat{X}_{[K_a]} \dmat{X}^H_{[K_a]}.
$$

Let us denote two random variables $\chi_1$ and $\chi_2$ as follows
\begin{eqnarray*}
\chi_1 &=& \normf{\dmat{P}_{[K_a+t]\backslash [t]}\rmat{Y}} - \normf{\dmat{P}_{[K_a]}\rmat{Y}} \\
&=& \sum_{l=1}^{L}\rvec{y}_{l}^H\left(
\dmat{P}_{[K_a+t]\backslash [t]} - \dmat{P}_{[K_a]} \right)\mathbf{y}_{l}
\end{eqnarray*}
and
\begin{eqnarray*}
\chi_2 &=& \alpha \normf{\dmat{P}_{[K_a] \backslash [t]}^\perp\rmat{Y}} - \normf{\dmat{P}_{[K_a]}^\perp\rmat{Y}} \\ 
&=& \sum_{l=1}^{L}\rvec{y}_{l}^H\left(
\alpha \dmat{P}_{[K_a] \backslash [t]}^\perp - \dmat{P}_{[K_a]}^\perp
\right)\rvec{y}_{l},
\end{eqnarray*}
where we used the following facts: $\normf{\dmat{A}} = \text{tr}{(\dmat{A}^H \dmat{A})}$, $\dmat{P}^H = \dmat{P}^2 = \dmat{P}$.

Note that 
\begin{eqnarray*}
\tilde{p}_{1,t} &=& \Pb{\chi_1 \geq 0, \ \chi_2\geq 0\} \;\middle|\; \rmat{X} = \dmat{X}} \\
&\leq& \mathbb{E}_{\rmat{Y}}\left[ \exp\left( \sum_{\ell=1}^{L}\rvec{y}_{l}^H \dmat{D} \rvec{y}_{l} \right) \right] \\
&=& \prod\limits_{l=1}^L \mathbb{E}_{\rvec{y}_l} \left[ \exp\left( \rvec{y}_{l}^H \dmat{D} \rvec{y}_{l} \right) \right] \\
&=& \determ{\dmat{I} - \dmat{D} \dmat{\Sigma}}^{-L},
\end{eqnarray*}
where the first inequality is due to Lemma~\ref{lemma:chernoff}, the second transition utilizes the fact that given $\dmat{X}_{[K_a]}$ the columns of $\rmat{Y}$ are i.i.d random vectors and finally the third equality is due to Lemma~\ref{lemma:qf_inv}.

Similarly, we deal with 
\begin{eqnarray*}
\Pb{ \mathcal{B}^c} = \binom{K_a}{t}\mathbb{E}_{\rmat{X}} \left[ \Pb{ \mathcal{B}_{[t]}^c \;\middle|\; \rmat{X}} \right]. 
\end{eqnarray*}
This concludes the proof.

\section{Complexity reduction via basis change}\label{sec:dem_reduction}

Note that evaluation of the bound from Theorem~\ref{thm} requires an optimization over parameters $u$, $v$, $\delta$ and $\alpha$. Each optimization step calculates the determinants $n\times n$ matrices ${\dmat{I} - \dmat{D} \dmat{\Sigma}}$ and ${\dmat{I} - \dmat{B} \dmat{\Sigma}}$, where the matrices $\dmat{D}$ and $\dmat{B}$ depend on the parameters $u$, $v$, $\delta$ and $\alpha$ (see \eqref{eq:D} and \eqref{eq:B}). This leads to a huge computational complexity. The complexity of this procedure can be significantly reduced. We apply a basis change such that these matrices become block-diagonal ones.

Let $\dset{T} = [K_a]$, $\dset{R} = [K_a+t]\backslash[t]$, $\dset{M} = [t]$, $\dset{F} = [K_a+t]\backslash[K_a]$ and $\dset{C} = [K_a]\backslash[t]$. Let us consider to the following basis
\[
\dmat{Q} = \left[ \dmat{Q}_c, \dmat{Q}_a, \dmat{Q}_m, \dmat{Q}_f \right], \quad \dmat{Q}^H\dmat{Q} = \dmat{I}_n
\]
where $\dmat{Q}_c$ is the ortonormal basis of $\Span{\dmat{X}_{\dset{C}}}$, $\dmat{Q}_m$ is the ortonormal basis of $\Span{\dmat{P}^\perp_{\dmat{B}_c} \dmat{X}_{\dset{M}}}$, $\dmat{Q}_f$ is the ortonormal basis of $\Span{\dmat{P}^\perp_{\dmat{B}_m} \dmat{P}^\perp_{\dmat{Q}_c} \dmat{X}_{\dset{F}}}$ and $\dmat{Q}_a$ appends $\dmat{Q}_f$ and $\dmat{Q}_m$ to the ortonormal basis $\left( \Span{\dmat{X}_{\dset{C}}} \right)^\perp$.

Clearly,
\begin{eqnarray*}
\determ{\dmat{I} - \dmat{D} \dmat{\Sigma}} &=& \determ{\dmat{Q}^{H} \left( \dmat{I} - \dmat{D} \dmat{\Sigma} \right) \dmat{Q}} \\
&=& \determ{\dmat{I} - (\dmat{Q}^{H} \dmat{D} \dmat{Q}) (\dmat{Q}^{H} \dmat{\Sigma} \dmat{Q})}.
\end{eqnarray*}

Let us start with $\dmat{Q}^{H} \dmat{D} \dmat{Q}$. Note that $\dmat{P}_\dset{T} = \dmat{P}_{\dset{C} \bigcup \dset{M}} = \dmat{P}_{\dset{C}} + \dmat{P}_{\dset{C}^\perp, \dset{M}}$, where $\dmat{P}_{ \dset{C}^\perp, \dset{M}}$ is a projection onto $\Span{\dmat{P}^\perp_{\dset{C}} \dmat{X}_{\dset{M}}}$. By analogy, $\dmat{P}_\dset{R} = \dmat{P}_{\dset{C} \bigcup \dset{F}} = \dmat{P}_{\dset{C}} + \dmat{P}_{\dset{C}^\perp, \dset{F}}$, where $\dmat{P}_{\dset{C}^\perp, \dset{F}}$ is a projection onto $\Span{\dmat{P}^\perp_{\dset{C}} \dmat{X}_{\dset{F}}}$. Let us rewrite the matrix $\dmat{D}$ as follows
\[
\dmat{D} = u \dmat{P}_{\dset{C}^\perp, \dset{F}} + (v-u) \dmat{P}_{\dset{C}^\perp, \dset{M}} + v (\alpha-1) \dmat{P}_{\dset{C}}^\perp.
\]

It is easy to check that
\[
\dmat{Q}^{H} \dmat{P}_{\dset{C}}^\perp \dmat{Q} = \begin{bmatrix}
\dmat{0}_{K_a-t} & \dmat{0} \\
\dmat{0} & \dmat{I}_{n-K_a+t}
\end{bmatrix},
\]
\[
\dmat{Q}^{H} \dmat{P}_{\dset{C}^\perp, \dset{M}} \dmat{Q} = \begin{bmatrix}
\dmat{0}_{n-2t} & \dmat{0} & \dmat{0} \\
\dmat{0} & \dmat{I}_{t} & \dmat{0} \\
\dmat{0} & \dmat{0} & \dmat{0}_t \\
\end{bmatrix},
\]
and
\[
\dmat{Q}^{H} \dmat{P}_{\dset{C}^\perp, \dset{F}} \dmat{Q} = \begin{bmatrix}
\dmat{0}_{n-2t} & \dmat{0} \\
\dmat{0} & \dmat{S}
\end{bmatrix},
\]
where $\dmat{S} = \left[ \dmat{Q}_m \dmat{Q}_f \right]^H  \dmat{P}_{\dset{C}^\perp, \dset{F}} \left[ \dmat{Q}_m \dmat{Q}_f \right]$.

Now consider $\dmat{B}^{H} \dmat{\Sigma} \dmat{B}$, we have
\[
\dmat{Q}^{H} \dmat{\Sigma} \dmat{Q} = \dmat{I}_n + \begin{bmatrix}
\dmat{T}_{1}     & \dmat{0}                & \dmat{T}_2    & \dmat{0}\\
\dmat{0}         & \dmat{0}_{n - K_a - t} & \dmat{0}      & \dmat{0}\\
\dmat{T}_3       & \dmat{0}                & \dmat{T}_4    & \dmat{0}\\
\dmat{0}         & \dmat{0}                & \dmat{0}      & \dmat{0}_t\\
\end{bmatrix},
\]
where $\dmat{T}_{1} = \dmat{Q}_c^{H} \dmat{X}_\dset{T} \dmat{X}^H_\dset{T} \dmat{Q}_c$, $\dmat{T}_{2} = \dmat{Q}_m^{H} \dmat{X}_\dset{M} \dmat{X}^H_\dset{M} \dmat{Q}_c$, $\dmat{T}_{3} = \dmat{Q}_c^{H} \dmat{X}_\dset{M} \dmat{X}^H_\dset{M} \dmat{Q}_m$ and $\dmat{T}_{4} = \dmat{Q}_m^{H} \dmat{X}_\dset{M} \dmat{X}^H_\dset{M} \dmat{Q}_m$. Here we used the following facts: (a) $\dmat{X}_\dset{T} \dmat{X}^H_\dset{T} = \dmat{X}_\dset{C} \dmat{X}^H_\dset{C} + \dmat{X}_\dset{M} \dmat{X}^H_\dset{M}$; (b) the columns of $\dmat{X}_\dset{C}$ are orthogonal to $\dmat{Q}_a$, $\dmat{Q}_m$ and $\dmat{Q}_f$; (c) the columns of $\dmat{X}_\dset{M}$ are orthogonal to $\dmat{Q}_a$ and $\dmat{Q}_f$.

Finally, we have
\[
\dmat{I} - \dmat{Q}^{H} \dmat{D} \dmat{Q} = 
\begin{bmatrix}
\dmat{I}_{K_a-t}   & \dmat{0}                & \dmat{0}  \\
\ldots             & (1-v(\alpha-1))\dmat{I}_{n - K_a - t} & \dmat{0}  \\
\vdots             & \ldots                  & \dmat{\hat{D}}
\end{bmatrix},
\]
where
\begin{eqnarray*}
\dmat{\hat{D}} &=& (1-v(\alpha-1))\dmat{I}_{2t} - u \dmat{S} - (v-u)\begin{bmatrix}
\dmat{I}_{t} & \dmat{0} \\
\dmat{0} & \dmat{0}_t \end{bmatrix} \\
&-& \left( (v\alpha-u) \dmat{I}_{2t} + u \dmat{S} \right) \begin{bmatrix}
\dmat{T}_{4} & \dmat{0} \\
\dmat{0} & \dmat{0}_t \end{bmatrix}
\end{eqnarray*}

Finally, we note that $\determ{\dmat{I} - \dmat{D} \dmat{\Sigma}} = (1-v(\alpha-1))^{n - K_a - t} \determ{\dmat{\hat{D}}}$ and the matrix $\dmat{\hat{D}}$ is of size $2t \times 2t$. 

\begin{remark}
In what follows we calculate the bound by sampling of the codewords $\dmat{X}_{[K_a+t]}$. We mention that it is sufficient to compute matrices $\dmat{S}$ and $\dmat{T}_4$ once per sample.  
\end{remark}

The case of $\determ{\dmat{I} - \dmat{B} \dmat{\Sigma}}$ is much easier. We note that (a) $\dmat{P}^\perp_{\dset{T}} \dmat{\Sigma} = \dmat{\Sigma} \dmat{P}^\perp_{\dset{T}} = \dmat{P}^\perp_{\dset{T}}$ and (b) $\dmat{P}^\perp_{\dset{T}} \dmat{P}^\perp_{\dset{C}} = \dmat{P}^\perp_{\dset{C}} \dmat{P}^\perp_{\dset{T}} = \dmat{P}^\perp_{\dset{T}}$, Thus,
\[
(\dmat{P}^\perp_{\dset{C}} \dmat{\Sigma}) (\dmat{P}^\perp_{\dset{T}} \dmat{\Sigma}) = \dmat{P}^\perp_{\dset{T}} = (\dmat{P}^\perp_{\dset{T}} \dmat{\Sigma}) (\dmat{P}^\perp_{\dset{C}} \dmat{\Sigma}).
\]

So the matrices $\dmat{P}^\perp_{\dset{C}} \dmat{\Sigma}$ and $\dmat{P}^\perp_{\dset{T}} \dmat{\Sigma}$ commute and can be diagonalized simultaneously.

\section{Numerical results}\label{sec:numres}

\begin{figure}
\centering
\includegraphics{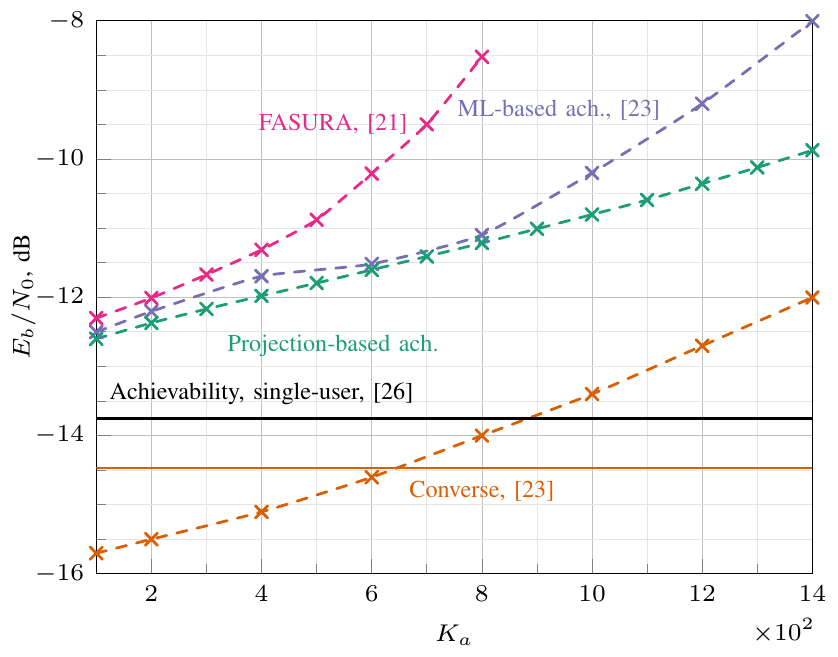}
\caption{Same-codebook achievability bound for no-CSI setting. Frame length $n=3200$, the number of information bits $k=100$. Base station is equipped with $L=50$ antennas, $P_e = 0.025$. Single-user converse is marked by solid line, multi-user converse is marked by dashed line.\label{fig:numerical_results}}
\end{figure}

In this section, we provide a numerical results for the same codebook setup~\cite{noPoorUnsourcedML2023}. Result are presented in~\Fig{fig:numerical_results}. The number of active user varies within $K_a \leq 1400$. For each $K_a$, we have evaluated $p_{1,t}$ and $p_{2,t}$ from~\eqref{eq:p1t} and~\eqref{eq:p2t} for all $t \in \left[K_a\right]$. We also note that small values of $t$ define the PUPE for this range of active users count. We have evaluated the expectation in~\eqref{eq:p1t} and~\eqref{eq:p2t} using $5000$ samples. For the projection-based achievability bound (shown by green line), the energy efficiency is approximately the same for $K_a \leq 200$. Then, as the number of active users grows, there is an energy efficiency gap compared to the ML-based bound (shown by blue line). As a reference, we have also added a converse bound (including multi-user converse from~\cite{noPoorUnsourcedML2023} and a single-user converse from~\cite{Polyanskiy2013SIMO}, marked by dashed and solid line respectively), as well as the singe-user achievability bound from~\cite{Polyanskiy2013SIMO}. One can observe a gap of $\approx 3$ dB between the achievability and converse bounds. With a further $K_a$ increase, this gap becomes smaller, and when $K_a$ approaches $n$, projection-based achievability bound shows worse results. To show this, we compare the ML-based and projection-based achievability for $k=100$, $n=1000$, $L=64$ and $P_e = 10^{-3}$. This scenario has been taken from~\cite{Poor2022ML}. We also find that FASURA~\cite{FASURA}, a state-of-the-art practical scheme, demonstrates the energy efficiency very close to both ML- and projection-based bounds for $K_a \leq 400$.

As a further research, we plan to find a theoretical methods of evaluating the expectations in~\eqref{eq:p1t} and~\eqref{eq:p2t} and evaluating the number of samples required to find these expectations reliably.

\begin{figure}
\centering
\includegraphics{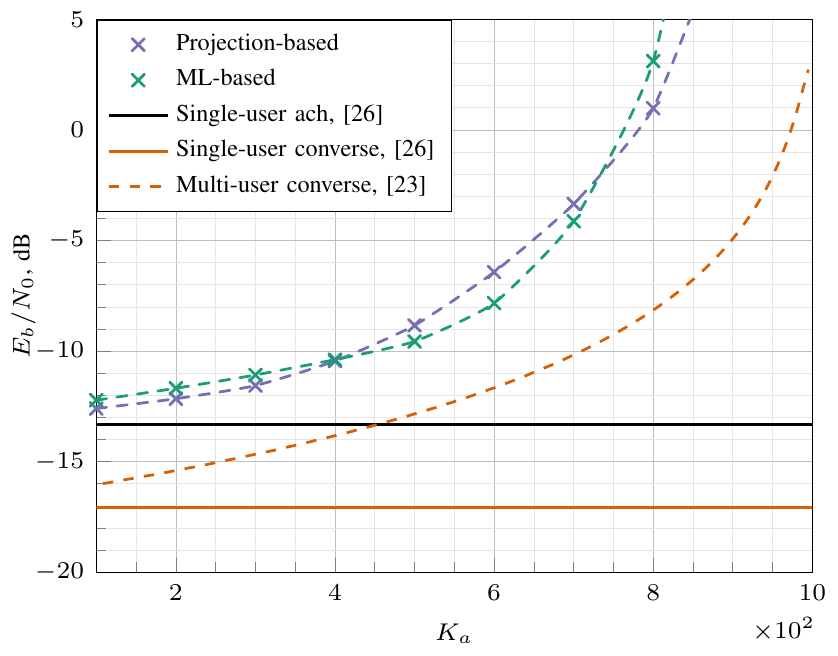}
\caption{Same-codebook achievability bound for no-CSI setting. Frame length $n=1000$, the number of information bits $k=100$. Base station is equipped with $L=64$ antennas, $P_e = 10^{-3}$.}
\end{figure}

\bibliographystyle{IEEEtran}
\bibliography{main}

\begin{thebibliography}{10}
\providecommand{\url}[1]{#1}
\csname url@samestyle\endcsname
\providecommand{\newblock}{\relax}
\providecommand{\bibinfo}[2]{#2}
\providecommand{\BIBentrySTDinterwordspacing}{\spaceskip=0pt\relax}
\providecommand{\BIBentryALTinterwordstretchfactor}{4}
\providecommand{\BIBentryALTinterwordspacing}{\spaceskip=\fontdimen2\font plus
\BIBentryALTinterwordstretchfactor\fontdimen3\font minus
  \fontdimen4\font\relax}
\providecommand{\BIBforeignlanguage}[2]{{%
\expandafter\ifx\csname l@#1\endcsname\relax
\typeout{** WARNING: IEEEtran.bst: No hyphenation pattern has been}%
\typeout{** loaded for the language `#1'. Using the pattern for}%
\typeout{** the default language instead.}%
\else
\language=\csname l@#1\endcsname
\fi
#2}}
\providecommand{\BIBdecl}{\relax}
\BIBdecl

\bibitem{polyanskiy2017perspective}
Y.~Polyanskiy, ``{A perspective on massive random-access},'' in
  \emph{Information Theory (ISIT), 2017 IEEE International Symposium on}.\hskip
  1em plus 0.5em minus 0.4em\relax IEEE, 2017, pp. 2523--2527.

\bibitem{ZPT-isit19}
I.~Zadik, Y.~Polyanskiy, and C.~Thrampoulidis, ``{Improved bounds on Gaussian
  MAC and sparse regression via Gaussian inequalities},'' in \emph{2019 IEEE
  International Symposium on Information Theory (ISIT)}.\hskip 1em plus 0.5em
  minus 0.4em\relax IEEE, 2019.

\bibitem{achRandUser2023}
K.-H. Ngo, A.~Lancho, G.~Durisi, and A.~Graell~i Amat, ``Unsourced multiple
  access with random user activity,'' \emph{IEEE Transactions on Information
  Theory}, pp. 1--1, 2023.

\bibitem{vem2017user}
A.~Vem, K.~R. Narayanan, J.~Cheng, and J.-F. Chamberland, ``{A user-independent
  serial interference cancellation based coding scheme for the unsourced random
  access Gaussian channel},'' in \emph{proc. IEEE Information Theory Workshop
  (ITW)}.\hskip 1em plus 0.5em minus 0.4em\relax IEEE, 2017, pp. 121--125.

\bibitem{Marshakov2019Polar}
E.~{Marshakov}, G.~{Balitskiy}, K.~{Andreev}, and A.~{Frolov}, ``{A Polar Code
  Based Unsourced Random Access for the Gaussian MAC},'' in \emph{proc. IEEE
  90th Vehicular Technology Conference (VTC2019-Fall)}, Sep. 2019, pp. 1--5.

\bibitem{codedCS2020}
V.~K. Amalladinne, J.-F. Chamberland, and K.~R. Narayanan, ``{A Coded
  Compressed Sensing Scheme for Unsourced Multiple Access},'' \emph{IEEE
  Transactions on Information Theory}, vol.~66, no.~10, pp. 6509--6533, 2020.

\bibitem{fengler2020sparcs}
A.~Fengler, P.~Jung, and G.~Caire, ``{SPARCs for Unsourced Random Access},''
  \emph{IEEE Transactions on Information Theory}, vol.~67, no.~10, pp.
  6894--6915, 2021.

\bibitem{AmalladinneAMP2021}
V.~K. Amalladinne, A.~K. Pradhan, C.~Rush, J.-F. Chamberland, and K.~R.
  Narayanan, ``{Unsourced Random Access With Coded Compressed Sensing:
  Integrating AMP and Belief Propagation},'' \emph{IEEE Transactions on
  Information Theory}, vol.~68, no.~4, pp. 2384--2409, 2022.

\bibitem{AmalladinneJoint2023}
A.~K. Pradhan, V.~K. Amalladinne, A.~Vem, K.~R. Narayanan, and J.-F.
  Chamberland, ``Sparse idma: A joint graph-based coding scheme for unsourced
  random access,'' \emph{IEEE Transactions on Communications}, vol.~70, no.~11,
  pp. 7124--7133, 2022.

\bibitem{Duman2021}
M.~J. Ahmadi and T.~M. Duman, ``{Random Spreading for Unsourced MAC With Power
  Diversity},'' \emph{IEEE Communications Letters}, vol.~25, no.~12, pp.
  3995--3999, 2021.

\bibitem{PradhanLDPCSC2021}
A.~K. Pradhan, V.~K. Amalladinne, K.~R. Narayanan, and J.-F. Chamberland,
  ``{LDPC Codes with Soft Interference Cancellation for Uncoordinated Unsourced
  Multiple Access},'' in \emph{ICC 2021 - IEEE International Conference on
  Communications}, 2021, pp. 1--6.

\bibitem{Kowshik2021TIT}
S.~S. Kowshik and Y.~Polyanskiy, ``{Fundamental Limits of Many-User MAC With
  Finite Payloads and Fading},'' \emph{IEEE Transactions on Information
  Theory}, vol.~67, no.~9, pp. 5853--5884, 2021.

\bibitem{Kowshik2020TCOM}
S.~S. {Kowshik}, K.~{Andreev}, A.~{Frolov}, and Y.~{Polyanskiy}, ``{Energy
  Efficient Coded Random Access for the Wireless Uplink},'' \emph{IEEE
  Transactions on Communications}, vol.~68, no.~8, pp. 4694--4708, 2020.

\bibitem{Truhachev2022}
E.~Nassaji, M.~Bashir, and D.~Truhachev, ``{Unsourced Random Access Over Fading
  Channels via Data Repetition, Permutation, and Scrambling},'' \emph{IEEE
  Transactions on Communications}, vol.~70, no.~2, pp. 1029--1042, 2022.

\bibitem{Andreev2022}
K.~Andreev, P.~Rybin, and A.~Frolov, ``{Coded Compressed Sensing With List
  Recoverable Codes for the Unsourced Random Access},'' \emph{IEEE Transactions
  on Communications}, vol.~70, no.~12, pp. 7886--7898, 2022.

\bibitem{Fengler2021}
A.~Fengler, S.~Haghighatshoar, P.~Jung, and G.~Caire, ``{Non-Bayesian Activity
  Detection, Large-Scale Fading Coefficient Estimation, and Unsourced Random
  Access With a Massive MIMO Receiver},'' \emph{IEEE Transactions on
  Information Theory}, vol.~67, no.~5, pp. 2925--2951, 2021.

\bibitem{Fengler2022}
A.~Fengler, O.~Musa, P.~Jung, and G.~Caire, ``{Pilot-Based Unsourced Random
  Access With a Massive MIMO Receiver, Interference Cancellation, and Power
  Control},'' \emph{IEEE Journal on Selected Areas in Communications}, vol.~40,
  no.~5, pp. 1522--1534, 2022.

\bibitem{Duman2022}
M.~J. Ahmadi and T.~M. Duman, ``{Unsourced Random Access with a Massive MIMO
  Receiver Using Multiple Stages of Orthogonal Pilots},'' in \emph{2022 IEEE
  International Symposium on Information Theory (ISIT)}, 2022, pp. 2880--2885.

\bibitem{AmalladinneMIMO2021}
\BIBentryALTinterwordspacing
V.~K. Amalladinne, J.-F. Chamberland, and K.~R. Narayanan, ``{Coded Compressed
  Sensing with Successive Cancellation List Decoding for Unsourced Random
  Access with Massive MIMO},'' 2021. [Online]. Available:
  \url{https://arxiv.org/abs/2105.02185}
\BIBentrySTDinterwordspacing

\bibitem{Shyianov2020}
V.~Shyianov, F.~Bellili, A.~Mezghani, and E.~Hossain, ``{Massive Unsourced
  Random Access Based on Uncoupled Compressive Sensing: Another Blessing of
  Massive MIMO},'' \emph{IEEE Journal on Selected Areas in Communications},
  vol.~39, no.~3, pp. 820--834, 2021.

\bibitem{FASURA}
M.~Gkagkos, K.~R. Narayanan, J.-F. Chamberland, and C.~N. Georghiades,
  ``{FASURA: A Scheme for Quasi-Static Massive MIMO Unsourced Random Access
  Channels},'' in \emph{2022 IEEE 23rd International Workshop on Signal
  Processing Advances in Wireless Communication (SPAWC)}, 2022, pp. 1--5.

\bibitem{Poor2022ML}
J.~Gao, Y.~Wu, S.~Shao, W.~Yang, and H.~Vincent~Poor, ``{Energy Efficiency of
  Massive Random Access in MIMO Quasi-Static Rayleigh Fading Channels with
  Finite Blocklength},'' \emph{IEEE Transactions on Information Theory}, pp.
  1--1, 2022.

\bibitem{noPoorUnsourcedML2023}
\BIBentryALTinterwordspacing
J.~Gao, Y.~Wu, T.~Li, and W.~Zhang, ``{Energy Efficiency of MIMO Massive
  Unsourced Random Access with Finite Blocklength},'' 2023. [Online].
  Available: \url{https://arxiv.org/abs/2302.02048}
\BIBentrySTDinterwordspacing

\bibitem{cover2012elements}
T.~M. Cover and J.~A. Thomas, \emph{{Elements of information theory}}.\hskip
  1em plus 0.5em minus 0.4em\relax John Wiley \& Sons, 2012.

\bibitem{mathai1992quadratic}
A.~Mathai and S.~Provost, \emph{{Quadratic Forms in Random Variables}}, ser.
  Statistics: A Series of Textbooks and Monographs.\hskip 1em plus 0.5em minus
  0.4em\relax Taylor \& Francis, 1992.

\bibitem{Polyanskiy2013SIMO}
W.~Yang, G.~Durisi, T.~Koch, and Y.~Polyanskiy, ``{Quasi-static SIMO fading
  channels at finite blocklength},'' in \emph{2013 IEEE International Symposium
  on Information Theory}, 2013, pp. 1531--1535.

\end{thebibliography}
\appendices
\end{document}